\documentclass[useAMS,usenatbib]{mn2e}

\usepackage{graphicx,amssymb,amsmath,times}

\title[Gamma-ray emission from PWNe]{Gamma-ray emission from PWNe interacting with molecular clouds}

\author[H. Bartko and W. Bednarek]{H. Bartko$^{1}$
and W. Bednarek$^{2}$ \\
$^{1}$Max-Planck-Institut f\"ur Physik, D-80805 M\"unchen, Germany, hbartko@mppmu.mpg.de\\
$^{2}$University of \L\'od\'z, PL-90236 Lodz, Poland, bednar@fizwe4.fic.uni.lodz.pl}
\begin{document}

\date{Accepted;  Received; in original form 1988 October 11}

\pagerange{\pageref{firstpage}--\pageref{lastpage}} \pubyear{200}

\maketitle

\label{firstpage}

\begin{abstract}
We consider a situation in which a pulsar is formed inside or close to a high density region of a molecular cloud.
Right after birth, the pulsar was very active and accelerated hadrons and leptons to very high energies. Hadrons diffuse through the supernova remnant (SNR) and some of them are trapped in the nearby cloud interacting with the matter. 
We extend a recent time-dependent model for the $\gamma$-radiation of 
pulsar wind nebulae (PWNe) to describe this more complicated astrophysical scenario.
The example calculations have been performed for two objects, IC443 and W41, which have recently been discovered as sources of TeV $\gamma$-rays. 
In this model the low energy TeV emission should be correlated with the birth place of the pulsar and the region of dense soft radiation rather than with its present position, provided that the injection rate of relativistic particles into the PWNa has been much more efficient at early times. The high energy TeV emission should be correlated with the location of dense clouds which were able to capture high energy hadrons due to their strong magnetic fields.

\end{abstract}



\section{Introduction}

A few recently discovered TeV $\gamma$-ray sources seem to be clearly related to pulsar wind nebulae (PWNe)~\citep{ah06a}. 
However, in some cases the TeV source is clearly displaced from the present position of the pulsar. This can be explained { either} by the fact that while the pulsar was very young it injected particles and then it moved significantly during its lifetime of $\sim$10$^4$ years { or by the displacement of the low energy pairs from their place of origin due to the interaction of the PWNa with the reverse shock of the supernova remnant (SNR) or by a back flow in the bow shock nebulae (e.g.~\citet{Blondin01}). All these explanations are likely and they cannot be excluded at the present stage of knowledge.} Here, we consider { the first scenario} in more detail { applying it to} two such example TeV $\gamma$-ray sources with evidences of associated energetic pulsars: IC~443 (MAGIC J0616+225,~\citet{al07}, see also \citet{Humensky2007}) and W41 (HESS J1834-087,~\citet{ah06a} and~\citet{al06}).

In the case of IC~443
the VHE $\gamma$-ray source is located in the direction of a dense molecular cloud which probably lies in front of the parent asymmetric SNR. Moreover, an X-ray nebula, G189.22+2.90, has been discovered near the shell of IC443 by recent {\it XMM-Newton} and {\it Chandra} observations~\citep{bb01,ol01}. It is interpreted as being due to the presence of an energetic Vela-type pulsar. The X-ray PWNa is displaced from the location of the TeV source by $\sim 20'$,~\citep{al07}. We suggest that this pulsar can also be responsible for the observed TeV $\gamma$-ray source. The $\gamma$-rays can be produced by particles accelerated inside the SNR during the early age of the pulsar. In fact, a pulsar with a velocity of $\sim$250~km/s should change its position by $\sim 18'$ (at a distance of 1.5~kpc) by the time it reaches the age of $3\times 10^4$ years.


The VHE $\gamma$-ray source HESS J1834-087 looks extended with a size of $\sim 12'$~\citep{al06,ah06a}. It is positionally coincident with the shell-type SNR, G23.3-0.3 (W41), with an age estimate of $\sim 8\times 10^4$ years. A Vela-type pulsar, PSR J1833-827, with the period of 85 ms, is also possibly connected with this supernova~\citep{gj95}. Its characteristic age and distance are consistent with the age and distance of the SNR W41. However, the pulsar is displaced from the center of the SNR (and also from the position of the TeV source) by $\sim 24'$. Such positional disagreement might again be explained by the movement of the pulsar during $\sim 10^5$ years with a velocity of $\sim 250$ km/s. The pulsar seems to be located in the direction of one of the extensions of the TeV $\gamma$-ray source~\citep{ah06a}. 

In this paper we extend the hybrid (leptonic-hadronic) time dependent radiation model for PWNe~(Bednarek \& Bartosik 2003, hereafter BB 03) for the more complicated case of fast-moving pulsars in the vicinity of dense molecular clouds.

%
%

\section{A model for $\gamma$-ray production}

We consider a scenario in which an energetic, fast-moving pulsar is created in a supernova explosion close to a dense cloud. The evolution of the Pulsar Wind Nebula, created by such a pulsar, can differ significantly from the case of an isolated PWNa of the Crab type. The higher density of the surrounding medium results in a slower expansion of the SNR and lower adiabatic losses of particles. The high density matter can provide additional soft radiation and matter targets for the particles accelerated in the vicinity of the pulsar. Moreover, the pulsar, changing position in time, injects relativistic particles (leptons and hadrons) in different places. 
%
%



The \citet{bb03} { radiation} model is a time-dependent, but one-space-dimensional model for isolated PWNe. { In order to consider in detail the time-dependent radiation processes inside the nebula \citet{bb03} consider a rather simple model for the expansion of the PWNa and the SNR based on the early models proposed in the literature (e.g.~\citep{ostriker1971,pacini1973,rees1974,reynolds1984}). Note however, that it is at present well known that the evolution of such systems is quite complicated (as considered by e.g.~\citep{Blondin01,vanderswaluw2001, bucciantini2003}). At the present stage we are not able to take into account the complicated evolution of the nebula envisaged in these papers. It can show significant differences in the case of nebulae with an age larger than a few thousand years. Therefore, we limit our considerations to a simple evolution model of the nebula and concentrate on a detailed description of the complicated time-dependent radiation processes inside the nebula. As we show farther, our model considers the acceleration of leptons at the relatively earlier stage of the PWN evolution. Therefore, the exact evolution of the nebula at later stages does not have a very important effect on the radiation processes. Still, it might have an effect on the morphology of the $\gamma$-ray source.
    
Our model applies the hypothesis of Arons and collaborators (see e.g. \citet{ho92,ga94}) according to which leptons gain energy by being resonantly scattered off the Alfven waves generated above the pulsar wind shock by hadrons injected by the pulsar.}
The model assumes that most of the rotational energy loss rate of the pulsar is taken by relativistic hadrons which can gain energy corresponding to approximately $20\%$ of the maximum potential drop through the pulsar magnetosphere. Therefore, hadrons are injected mono-energetically. According to \citet{ho92} leptons gain a power-law spectrum with an index not far from two. In our model we assume this power-law index to be independent of time.

These leptons and hadrons diffuse inside the SNR and loose energy in adiabatic and radiation processes. The diffusion of leptons is much slower than the one of the hadrons due to huge radiation losses on synchrotron, bremsstrahlung and IC processes. Therefore, hadrons are able to partially escape from the SNR, contrary to leptons.
{ Note, however, that we apply a rather simple model for the diffusion of hadrons from the nebula~\citep{bp02}. This model does not take into account the Rayleigh-Taylor instabilities on the border separating the PWNa from the SNR. Due to these instabilities, the escape of hadrons from the nebula may become more efficient and the target for relativistic hadrons which are still inside the PWNa may increase. These two effects have an opposite influence on the $\gamma$-ray spectrum produced by hadrons inside and outside the PWNa and are not taken into account in our model}.


Due to the evolution of the SNR in time (radius, expansion velocity, density of matter, magnetic field strength, see Eqs.~(1-7) in~\citet{bb03}) the above mentioned processes are time-dependent in a non-trivial way. The evolution of the equilibrium spectra of leptons is described by Eqs. (11-12) of ~\citet{bb03} and the one of hadrons by Eqs. (10-11) of~\citet{bb04}.
In the considered time-dependent scenario these equations cannot be solved analytically. We apply an iterative procedure, the time-step method, described in \citet{bb03}.


Based on the equilibrium spectra of particles we calculate the $\gamma$-ray production from the SNR in different radiation processes (bremsstrahlung and inverse Compton (IC) for leptons and interactions with matter for hadrons). Hadrons which escaped from the SNR can be partially captured by molecular clouds and also produce $\gamma$-rays. In the calculations of the $\gamma$-ray spectra for hadrons we apply the scaling break model for hadronic interactions~\citep{wd87} suitable for the considered energy range of hadrons.


In order to be able to apply this time-dependent model for $\gamma$-ray production from PWN to the considered astrophysical situation, it has to be extended by two issues: the fast motion of the pulsar and the vicinity of a dense molecular cloud.

The analysis of the radiation processes around pulsars is considerably complicated in the case of pulsars which are able to significantly change their later position with respect to their birth place. This situation obviously concerns the Vela-type pulsars which reach large velocities at birth and which are at later evolution stages still energetic enough to produce magneto-spheric $\gamma$-rays ($\gamma$-ray pulsars). In their case, the radiation processes described above can be additionally distributed along the path of the moving pulsar. In such a complicated case, the $\gamma$-ray source can be shifted with respect to the present position of the pulsar depending on the history of the injection of relativistic particles and the distribution of the target (soft radiation, matter) close to the path of the pulsar. 

In order to compute the spectrum and the morphology of the VHE $\gamma$-ray source a full three-space-dimensional and time-dependent treatment is necessary. Unfortunately, this is too difficult (requires consideration of three-dimensional radiation and diffusion processes) to be constructed in detail at present. We limit our calculations to a spherically symmetric case, which can give a correct estimate for the $\gamma$-ray spectrum and flux since the soft radiation fields (CMB, infrared and optical background photons in the Galactic plane or produced by the molecular cloud) do not depend on the movement of the pulsar. Note, that the IC scattering of the synchrotron radiation is negligible for old PWN. However, in such an approach we are not able to calculate the $\gamma$-ray source morphology.
In this spherically symmetric model, we 
can take into account the additional soft radiation target, created by the emission from the molecular cloud, but we cannot take into account the possible influence of the cloud on the spherical geometry of the expanding SNR and the related PWNa. Therefore, this model can give an approximately correct description of the considered picture only in the case that the cloud is not too close to the expanding SNR.


There are arguments (see Sect. 3 in \citet{bb03}) that the injection rate of particles into the nebula can be limited to a specific period after the pulsar formation. The processes occurring close to the pulsar surface or inside its inner magnetosphere are only active i.e. during the first $10^4-10^5$ years after the pulsar formation, see e.g.~\citet{um95}. For example, nuclei, which are responsible for the acceleration of leptons, are only injected when the surface temperature of the neutron star is high enough.

In our calculations we apply the basic parameters of the exploding supernova as used in our earlier modeling of the Crab Nebula~\citep{bb03}, i.e the initial period of the pulsar is 10 ms, the initial velocity of the supernova shock is 2000 km s$^{-1}$, and the mass of the nebula is 4 M$_\odot$.  
We assume that the supernova exploded in a medium with a density of $\sim 20$
part. cm$^{-3}$ (Dame et al.~1986) and a magnetic field of $10^{-5}$ G. Such a large density is supported by the presence of dense clouds around the SNR.
Following Arons and collaborators (e.g.~\citet{ga94}), we assume that hadrons take $95\%$ of the rotational energy loss of the pulsar. Only $5\%$ of the energy of hadrons is transferred to leptons. 

The infrared soft radiation field is expected to be considerably stronger inside the SNR due to the emission from the dust heated by the supernova shock. Following other works, we apply the soft radiation field of the interstellar medium at the galactic disk as defined in~\citep{de97,ga98}. It is composed of the cosmic microwave background radiation (CMB) with a temperature of 2.7 K, the infrared background with an energy density two times larger than the CMB and a temperature of 25 K, the optical background defined by energy densities equal to the CMB and with characteristic temperatures between 5000 K and $10^4$ K. Moreover, we add the infrared radiation field with an energy density three times larger than the CMB and with a temperature of 45 K, which is supposed to be produced by dust inside the molecular cloud (based on the IRAS data \citep{mu86}).

\section{IC~443/MAGIC J0616+225}

\begin{figure*}[t]
\includegraphics[width=\columnwidth]{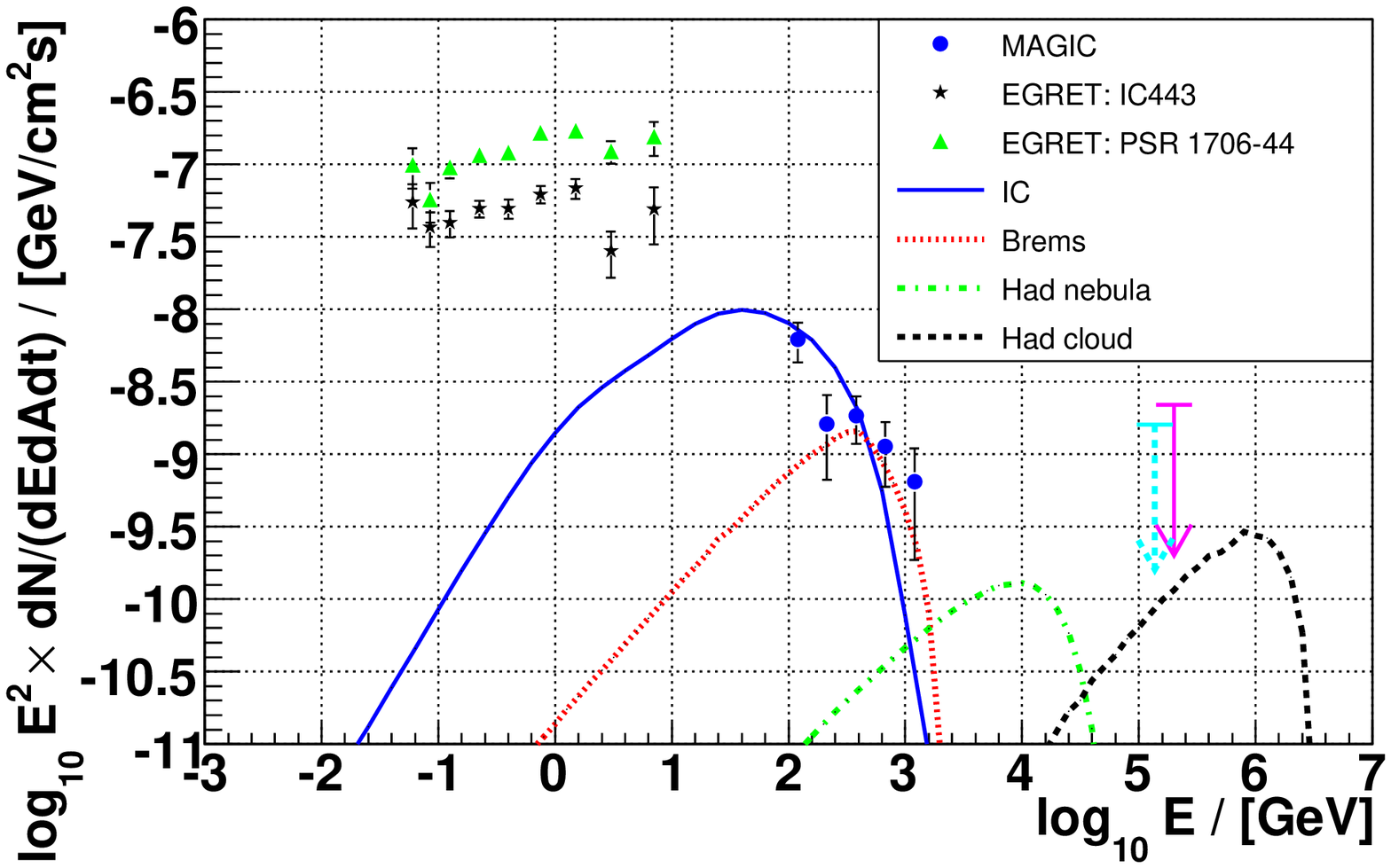}
\includegraphics[width=\columnwidth]{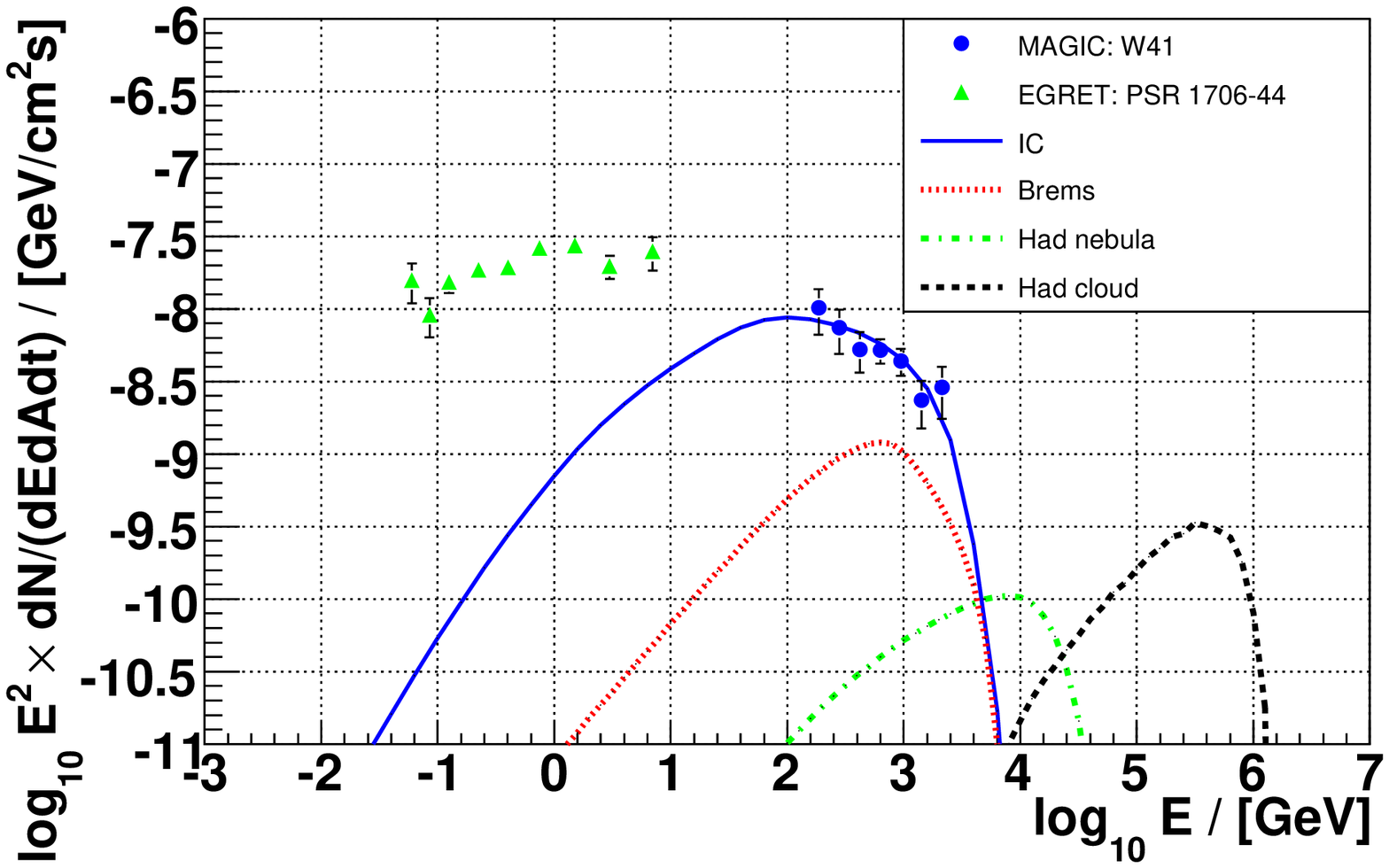}
\caption{\label{fig1} The $\gamma$-ray spectra (SED) produced by leptons in the
IC (blue full curve) and bremsstrahlung (red dotted curve) processes inside the SNRs 
IC~443 (left figure) and 
W41 (right figure).
The $\gamma$-ray spectra produced by hadrons 
inside the supernova remnant (green dot-dashed curve), and by hadrons which escaped 
from the supernova remnant and which were captured by nearby dense clouds 
(black dashed curves). We also show the $\gamma$-ray spectrum observed from the inner 
magnetosphere of the Vela-type pulsar (PSR 1706-44) by the EGRET telescope (green triangles), unscaled for IC~443 due to comparable distances and scaled to take into account the different distances of PSR 1706-44 and W41.
The EGRET data on IC~443 are portrayed by black stars. The MAGIC VHE $\gamma$-ray data are portrayed by blue points. The cyan and magenta arrows indicate 90$\%$ upper limits to the UHE $\gamma$-ray flux from the CASA-MIA and CYGNUS collaborations \citep{Allen1995,Borione1995}.
The parameters of the model are given in table \ref{tab1}. 
}
\end{figure*}

The parameters of the pulsar which creates the PWNa observed close to the SNR IC~443, G189.22+2.90, are unknown. But from comparison with other nebulae, \citet{ol01} argued that the pulsar has to be of a Vela-type with a present period of $\sim$145 ms, a surface  magnetic field of $\sim$3$\times 10^{12}$ G, and an age of $\sim$3$\times 10^4$ years, see also 
\citet{bb01}. The displacement between the present location of the 
PWNa and the center of the SNR can be explained by the motion of the pulsar with a velocity of $\sim$250 km/s taking into account its estimated age. 
IC~443 is an asymmetric shell-type SNR with a diameter of 45' 
at a distance of about 1.5~kpc \citep{fe84,cl97}.
It is a prominent X-ray source, see e.g. \citet{Troja2006,ol01,Gaensler2006}.
The EGRET has detected a $\gamma$-ray source above 100 MeV in the IC~443 SNR, 3EG J0617+2238 \citep{Eposito1996,EGRET}.
There is a large amount of molecular mass ($\lesssim 10^{4}$
M$_{\odot}$) consistent with the distance to the SNR IC~443
\citep{Torres2002}. 
The highest CO intensity detected is directly superimposed on
the central position of the VHE $\gamma$-ray source MAGIC J0616+225 \citep{al07}. 

The pulsar with the parameters mentioned above resembles the $\gamma$-ray pulsars in Vela and PSR 1706-44. 
In fact, when we compare the $\gamma$-ray spectrum observed from e.g. PSR 1706-44
with the spectrum of the EGRET source 3EG J0617+2238, which is observed in
the direction of IC~443, we get good consistency of the spectral shapes and fluxes (within a factor of two) of these two sources (see Fig.~1). Note, that the distances of the pulsars have significant errors due to non-exactly known parameters of the interstellar space.
Therefore, we propose that, in spite of only marginal directional agreement, the EGRET source 3EG J0617+2238 is due to the pulsar responsible for the PWNa observed close to IC~443.

To calculate the $\gamma$-ray spectra produced by leptons in the IC and bremsstrahlung processes we used the above described model with parameters of the supernova remnant, the pulsar, and surrounding medium mentioned at the end of section 2 and in Tab. 1. We investigated the dependence of the flux level and spectra on the duration of the lepton injection phase and their spectral index, keeping all other parameters fixed. We found that the MAGIC data can be fitted well by the values of these parameters reported in Tab. 1.
The difference between the present location of the PWNa
and the position of the MAGIC TeV $\gamma$-ray source can be explained by 
assuming that leptons are accelerated only during the early
phase of the PWNa.

The population of high energy electrons, which produces the VHE $\gamma$-rays by IC scattering and bremsstrahlung, also produces synchrotron radiation in the magnetic field of the SNR. However, these old electrons do not have a sufficiently high enough energy to produce synchrotron X-rays. The characteristic energy of synchrotron photons can be estimated by $\epsilon_{\mathrm{synch}} \approx m_e \gamma_e^2 B/(4 \cdot 10^{13}\mathrm{G})$. Therefore, 10 TeV electrons would produce synchrotron emission  with a characteristic energy of only about 50 eV.

We calculate the $\gamma$-ray spectrum from the decay of $\pi^{0}$s produced by hadrons accelerated by the PWNa and interacting with the matter inside this nebula. However, since the nebula is relatively old, most of the hadrons accelerated during the early stage of the nebula have already escaped from it. Therefore, the level of the $\gamma$-ray emission produced by hadrons which are still inside the SNR is relatively low (see Fig. 1).

We also estimate the $\gamma$-ray fluxes produced by those hadrons, which escaped
from the PWNa in the past activity stage of the pulsar. It is very difficult
to reliably estimate the part of hadrons which are captured inside dense
clouds, since this process depends on the geometry of the cloud and the nebula, their relative distance, and above 
all on the geometry and distribution of the magnetic field in the region surrounding the PWNa.
This accumulation of hadrons may (or may not) also depend on the energy of the accelerated hadrons, depending on the details of the capturing process (e.g. whether it is mainly due to energy-dependent diffusion or due to advection from the wind from the pulsar). 
Since we are not able to take into account these complicated processes, it is assumed 
that a small part of all hadrons escaping from the PWNa is captured inside dense clouds.
The example calculations are performed for the parameters of the PWNa related to IC~443, 
assuming the presence of a nearby cloud with a density of $10^3$ particles cm$^{-3}$
and an efficiency of accumulation of hadrons equal to $10^{-5}$. 
It is clear that, even for such a very low capturing efficiency, these hadrons can produce
large fluxes of $\gamma$-rays in the energy range of $\sim$10-10$^3$ TeV. 
Note, that the upper limits to the $\gamma$-ray fluxes above 137/200~TeV by the Cygnus / CASA-MIA collaborations \citep{Allen1995,Borione1995} are more than one order of magnitude above our predicted flux. These upper limits constrain the capturing factor of hadrons in the molecular cloud to be less than $10^{-4}$. 

\begin{table} 
      \caption{Parameters of the model considered for IC 443 and W41} 
         \begin{tabular}{l l l} 
            \hline 
            \hline
                                   & IC 443              & W41\\ 

            \hline
 pulsar or PWNa                     & G189.22+2.90        &  PSR J1833-827  \\             \hline

pulsar surface magnetic field      & $3\times 10^{12}$ G & $1.2\times 10^{12}$ G \\ 
             \hline 
present pulsar period              & 145 ms              & 85 ms \\ 
                                   & (estimated)         & (observed) \\

             \hline 
age of the pulsar                  & $3\times 10^4$ yrs  & $8\times 10^4$ yrs \\
            \hline
duration of lepton injection phase & $10^4$ yrs          & $4\times 10^4$ yrs\\
            \hline 
spectral index of leptons          & 2.4                 &  2.4\\
            \hline             
density of surrounding medium      & 20 cm$^{-3}$        & 20 cm$^{-3}$ \\
            \hline
magnetic field of surrounding medium & $10^{-5}$~G         & $10^{-5}$~G \\ 
            \hline 
            \hline
         \end{tabular} 
         \label{tab1} 
     \end{table}

\section{W41/PSR J1833-827/HESS J1834-087}

The pulsar, PSR J1833-827, has been observed $24$' away from the center of the TeV $\gamma$-ray source HESS J1834-087. The pulsar has a period of 85 ms (belongs to the Vela-type pulsar population), has an estimated surface magnetic field of $\sim 10^{12}$ G, and lies at a distance of $\sim$4 kpc~\citep{gj95}. 
The characteristic age of the pulsar, $\sim 10^5$ years, and distance are consistent with the corresponding parameters of the SNR G23.3-0.3 (W41, i.e. an age of $8\times 10^4$ years~\citep{ti06}). The observed displacement between the present position of the PSR J1833-827 and the center of W41 can be easily explained by applying the above-mentioned age of the pulsar (and SNR) and assuming that the velocity of the pulsar is equal to $250$ km s$^{-1}$.
Recently, \citet{Landi2006} found a faint X-ray source within the area of HESS J1834-087/W41 in data from the Swift satellite and \citet{ti06} found an extended X-ray feature spatially coincident with the VHE $\gamma$-ray emission.
As in the case of IC~443, W41 is associated with a large molecular complex called "[23,78]''~\citep{da86} at a similar distance. 

We also apply the time-dependent model for the acceleration of hadrons and leptons in the vicinity of energetic pulsars for this object. The general parameters of the medium, the target photon fields with which particles interact,
the initial parameters of the supernova remnant and the pulsar are kept the same, as in the case of IC~443. 
We found that the MAGIC data can be fitted well by the same spectral index of leptons as in the case of IC~443, but by a four times larger duration of the lepton injection phase.
The results of calculations are also shown in Fig. 1.
In our interpretation, the differences between the $\gamma$-ray spectra produced by these two objects are mainly due to different values of the pulsar surface magnetic fields and the duration of the lepton injection phase. They determine the energy distribution and the power of the accelerated particles as well as their energy losses in case of leptons (synchrotron and indirect IC process) and the diffusion of hadrons inside the nebula.
Note, that the leptons produce $\gamma$-rays not at the present position of the pulsar, but closer to its birth position and the position of the molecular cloud due to the stronger infrared photon field of the cloud. 

We also calculate the $\gamma$-ray spectra from the decay of pions produced by hadrons, accelerated in the vicinity of the pulsar, which are
still inside the SNR, and those ones which escaped from the SNR and interact with dense clouds.

Note, that we scaled the EGRET data of PSR 1706-44 in the right plot of Fig.1 to account for the different distances of PSR 1706-44 and W41. In this way we predict the expected level of the flux in the EGRET energy range from the pulsar inside W41 which will be probed by the AGILE and GLAST telescopes.

\section{Conclusion}

{ We interpret the recently detected VHE $\gamma$-ray emission from two objects, IC 443 and W41, in terms of the time-dependent leptonic-hadronic model for the radiation processes in SNRs containing Vela-type pulsars. We modify the model~\citep{bb03} in order to take into account the presence of a nearby high density
region which can serve as a target for relativistic hadrons, escaping from the SNR,
and as an additional source of infrared photons upscattered to $\gamma$-ray energies by leptons inside the SNR. We show that} 
the higher density of the surrounding medium results in a slower expansion of the SNR and lower adiabatic losses of particles in respect to isolated SNRs. Moreover, the nearby high-density matter acts as a target for hadronic interactions and provides additional infrared radiation fields for more efficient leptonic interactions. All of this causes higher $\gamma$-ray fluxes.

We show that, in the case of IC 443 and W41, the observed $\gamma$-radiation can be explained by such SNR-PWN-pulsar systems for likely initial parameters of the pulsars and the SNRs and the surrounding medium. 
We suggest that the pulsar, the $\gamma$-ray sources at different energies, and the center of the parent SNR do not need to be co-spatial in the case of fast-moving pulsars close to a high-density region when the acceleration of leptons is much more efficient at the early age of the pulsar. $\gamma$-ray sources at energies below a few TeV are likely to appear in the place where the pulsars have been born and the $\gamma$-ray sources at higher energies should be related to the location of the massive nearby clouds. 
On the other hand, the GeV $\gamma$-ray source should be co-spatial to the present position of the pulsar.


\section*{Acknowledgements}

This work is supported by the German MPG and the Polish MNiI grant 1P03D01028.

\end{document}